\documentclass[11pt,a4paper]{article}
\pdfoutput=1
\usepackage{jcappub}
\usepackage{graphicx}
\usepackage{amsmath}   
\usepackage{amssymb}

\title{It is Feasible to Directly Measure Black Hole Masses in the First 
Galaxies}

\author[a]{Hamsa Padmanabhan,}

\author[b]{Abraham Loeb}

\affiliation[a]{Canadian Institute for Theoretical Astrophysics \\
60 St. George Street, Toronto, ON M5S 3H8, Canada}

\affiliation[b]{Astronomy department, Harvard University \\
60 Garden Street, Cambridge, MA 02138, USA}

\emailAdd{hamsa@cita.utoronto.ca}
\emailAdd{aloeb@cfa.harvard.edu}

\abstract{
In the local universe, black hole masses have been inferred from the observed 
increase in the velocities of stars at the centres of their host galaxies. So 
far, masses of supermassive black 
holes in the early universe have only been inferred indirectly, using 
relationships calibrated to their locally observed 
counterparts. Here, we use the latest observational constraints on the 
evolution 
of stellar masses in galaxies to predict,  {for the first time}, that the 
region 
of influence of a 
central supermassive black hole at the epochs where the first galaxies were 
formed is \textit{directly resolvable} by current and upcoming telescopes. { 
{We 
show that the existence of the black hole can be inferred from observations of 
the gas or stellar disc out to $> 0.5$ kpc from the host halo at redshifts $ z 
\gtrsim 6$.}} Such 
measurements will usher in a new era of discoveries unraveling the formation of 
the first supermassive black holes based on subarcsecond-scale spectroscopy 
with 
the 
$JWST$, $ALMA$,  and the $SKA$. 
The measured mass distribution of black holes will allow forecasting of the 
future detection of gravitational waves from the earliest black hole mergers.
}

\begin{document}
\maketitle
\flushbottom


\section{Introduction}

Massive black holes (with masses of about a million solar masses) are known to 
exist at the centres of galaxies in the nearby universe. Locally, the masses of 
such black holes have been measured from their gravitational radius of 
influence 
on the motion of stars and gas in their host galaxies. This method was 
particularly effective for the Milky Way \citep{ghez2005,ghez2008, genzel2000, 
schodel2003} and NGC 4258 \citep{miyoshi1995}. Local measurements imply a ratio 
of the central black hole mass to stellar mass $M_{*}$ of the host spheroid to 
be about $10^{-4} - 10^{-3}$ in the local universe \citep{kormendy2013, 
reines2015}. 
 
 Supermassive black holes with masses $M_{\rm BH}$ of more than a billion solar 
masses are inferred to exist out to a redshift $z = 7.54$, less than $\sim 700$ 
million years after the Big Bang \citep{banados2018}.  How the seeds for these 
black holes grew so early in the universe is still an open question 
\citep{inayoshi2019}. 
However, there is evidence indicating that at redshifts $z > 6$, the central 
black hole to stellar mass ratio had been more than order of magnitude or two 
higher \citep{decarli2018, venemans2016} than the locally observed value. 
Theoretical arguments supporting this evolution stem from the inferred black 
hole - bulge mass relation \citep{wyithe2002, croton2009} extrapolated to high 
redshifts, suggesting that $M_{\rm BH}$ grows as a steep function of the halo 
circular velocity $v_c$: $M_{\rm BH} \propto v_c^{\gamma}$ with $\gamma \sim 
5$. 
 
The evolution of the stellar mass $M_*$ in dark matter haloes of both early and 
present-day galaxies is fairly well-constrained. The latest observations from a 
wide range of surveys, including the Sloan Digital Sky Survey (SDSS), the PRIsm 
MUlti-object Survey (PRIMUS), UltraVISTA, the Cosmic Assembly Near-infrared 
Deep 
Extragalactic Legacy Survey (CANDELS), and the FourStar Galaxy Evolution Survey 
(ZFOURGE) all support \citep{behroozi2019} a well-defined $M_{*}(z)$ across the 
full redshift range, $0 < z < 10.5$. The data indicate that the average stellar 
mass in Milky-Way sized dark matter haloes evolves only by a factor of $\sim 
1.6$ over redshifts $z \sim 6$ to the present (see Fig. 9 of 
Ref. \citep{behroozi2019}). 

The above two results, taken together, imply a fascinating conclusion. Since 
the 
stellar mass evolves much more modestly than the black hole mass \citep[a trend 
also 
found in recent AGN observations from the 
\textit{Chandra}-COSMOS Legacy Survey; Ref.][]{suh2019}, the central black hole 
mass makes up a 
much larger fraction of the stellar mass at high redshifts (a fraction of $\sim 
0.3$ at redshift $z \sim 6$ compared to $10^{-4}$ at $z \sim 0$). This fact, 
{whose validity has been hinted at in the past \citep{davis2014,venemans2016}}, 
is well-supported by the recent 
$ALMA$\footnote{https://almascience.nrao.edu/about-alma/alma-basics} 
spectroscopic study 
\citep{decarli2018} of [CII]-detected quasars at $z > 6$ which indicates 
significantly more than two orders of magnitude evolution of the locally 
observed black hole - bulge mass relation \citep{reines2015}.\footnote{We 
also note that the recently discovered class of Obese Black Hole Galaxies 
\citep[OBGs;][]{natarajan2017} can have black hole - stellar mass ratios reaching as high as 
$M_{\rm BH}/M_{*} \sim 1$ at high redshifts.} An immediate 
implication of this finding is that the central black hole should have a 
measurable influence on the circular velocity profiles of galaxies at high 
redshift. A couple of caveats to this conclusion, however, are 
worth noting: 
it has been pointed out in the literature that the evolution in the black hole 
- 
bulge mass relation may be a selection effect \citep[e.g.,][]{lauer2007,
schulze2011, schulze2014, willott2017}, whereas some simulation papers do not 
find much 
evolution \citep[e.g.,][]{degraf2015, huang2018, marshall2019}. { {In the following section, we quantify the magnitude of this effect 
and its observational consequences for current and future generation 
experiments.}}

\begin{figure*}
\begin{center}
\hskip-0.5in \includegraphics[width = \textwidth]{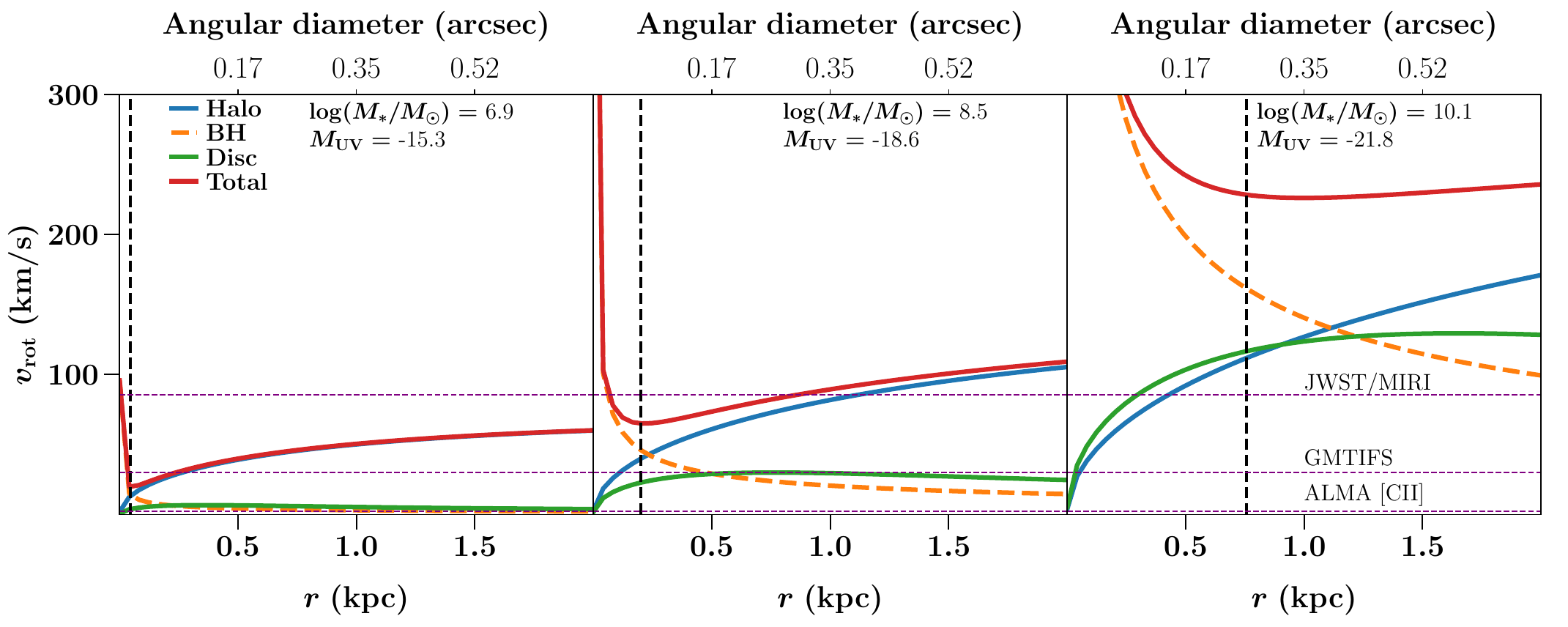}
\caption{Rotation curves around three typical dark matter haloes (with masses 
$M_{\rm halo} =  10^{10}, 10^{11}$ and $10^{12} M_{\odot}$ respectively from 
left to right) at $z = 6$. The halo is  assumed to follow a 
Navarro-Frenk-White \citep{nfw} profile with a concentration 
parameter \citep{diemer2018} $c = 3.5$, and contains a  central supermassive 
black hole, with mass $M_{\rm BH}$, and an exponential stellar disc with spin 
parameter $\lambda = 0.03$ (the inferred total stellar mass $M_*$ is also 
indicated,  {along with the corresponding UV magnitude, $M_{\rm UV}$ obtained 
from the relation of Ref. \citep{tacchella2018}}).  The circular velocities 
$v_{\rm rot}$ (in km/s) induced by the 
black hole (orange dashed lines), stellar disc (green solid lines) and dark 
matter halo (blue solid lines) are plotted along with the total circular 
velocity (red solid line). The central dominance of the black hole leads to a 
visible Keplerian correction in the otherwise flat rotation curve  within the 
inner $<1$ kpc. Vertical dashed lines in each panel indicate $R_{\inf}$, the 
radius of 
influence at which $M_{\rm BH} = M_{\rm halo}(R_{\rm inf}) + M_{\rm d}(R_{\rm 
inf})$. Horizontal dashed lines indicate the maximum spectral resolutions 
achievable by several current and upcoming facilities, such as the ALMA 
observing the [CII] 158 $\mu$m transition at $z \sim 6$, the GMT IFS and the 
JWST MIRI instrument. Top  horizontal axes show the angular diameter $\theta = 
[2r/D_A(z)]$ 
corresponding to the radial distance $r$ on the lower horizontal axis.}
\label{fig:rotvelcurves}
\end{center}
\end{figure*}

\section{Quantifying this effect}

To quantify this effect, we predict the rotation curves of typical Milky-Way 
sized haloes at $z \sim 6$, separating the dark matter,  stellar disc,  and 
black hole components. { {For the dark matter, we assume a Navarro-Frenk-White 
\citep{nfw} profile with the mass $M_{\rm halo}(r)$ within radius 
$r$ given by \citep{momaowhite}:
\begin{equation}
M_{\rm halo}(r) = 4 \pi \rho_{\rm crit} \delta_0 r_s^3 \left[ \ln(1 + cx) - 
\frac{cx}{1 + cx} \right]
\end{equation}
where $\rho_{\rm crit}$ is the critical density, and $c = r_{200}/r_s$ is the 
concentration parameter of the halo, expressed in terms of the limiting radius 
$r_{200}$ of the virialized halo and its scale radius $r_s$. The term $x$ 
denotes the radial distance expressed in terms of the limiting radius: $x = 
r/r_{200}$. The $\delta_0$ is the characteristic overdensity defined as:
\begin{equation}
\delta_0 = \frac{200}{3} \frac{c^3}{\ln(1+c) - c/(1+c)}
\end{equation}
The halo concentration is assumed 
to be $c = 3.5$ at $z = 6$, consistent with the findings of the recent 
physically motivated models \citep{diemer2018}. For the stellar component, we 
use an exponential disc profile, with the disc masses within a distance $r$ 
from 
the centre of the galaxy, $M_d(r)$ given by \citep{momaowhite}:
\begin{equation}
M_d(r) = M_d \left[1 - \left(1 + \frac{r}{R_d}\right) e^{-r/R_d}\right]
\end{equation}
with the disc scalelength, $R_d$ given by $R_d = \lambda r _{200} (j_d/m_d) 
(1/\sqrt{2})$
assuming a non-evolving specific angular momentum parameter 
($j_d/m_d \approx 1$) and spin parameter ($\lambda = 0.03$). The total disc 
mass 
$M_d$ is given by $M_d = m_d V_c^3/10 G H(z)$, where $V_c$ is the circular 
velocity of the halo and $H(z)$ is the Hubble constant at redshift $z$. The  
stellar mass is assumed to reside primarily in the disc, ($m_d = M_*/M_{\rm 
halo}$), with $M_*$ being the average stellar mass in the halo computed from 
the 
relation of Ref. \citep{behroozi2019}. The black hole is assumed to be a 
central 
point source with mass $M_{\rm BH}$ given by Ref. \citep{wyithe2002}.
}}

\section{Results}

Figure \ref{fig:rotvelcurves} shows the above three components at $z \sim 6$ 
for 
three representative dark matter halos with masses $M_{\rm halo} = 10^{10}, 
10^{11}$ and $10^{12} M_{\odot}$, respectively. The inferred total stellar 
masses $M_*$ from Ref. \citep{behroozi2019} are indicated on each panel, { {as 
well as the corresponding UV magnitudes, $M_{\rm UV}$ obtained from the 
observationally derived relation of Ref. \citep{tacchella2018}}}. Our 
predicted disc radii evolve as $(1+z)^{\beta}$ with $\beta \approx -1$ for this 
stellar mass range, consistent with the findings from the CANDELS 
\citep{kawamata2018} and HST/Hubble Frontier Fields 
\citep{shibuya2019,holwerda2015} galaxy observations at $z \sim 6-8$. The black 
hole masses in these galaxies predicted from Ref. \citep{wyithe2002} are 
$\log_{10} 
(M_{\rm BH}/M_{\odot}) = 6.3, 8.0, 9.7$, implying $M_{\rm BH}/M_{*} \approx 
0.3-0.4$. The figure shows rotational circular velocities, $v_{\rm rot, i} = (G 
M_i(r)/r)^{1/2}$  as a function of radial distance $r$, where $i$ denotes the 
disc, black hole or halo component. The total circular velocity, given by 
$v_{\rm tot}(r) = (\sum_i v_{\rm rot, i}(r)^2)^{1/2}$ is also plotted. A 
Keplerian correction to the rotation curve at the innermost $< 1$ kpc is 
clearly 
distinguishable for stellar masses of $10^9-10^{10} M_{\odot}$, a mass range 
for 
which sub-millimetre observations already exist in the literature 
\citep{shibuya2019}.  The radius of influence $R_{\rm inf}$ of the black hole, 
defined as the scale at which $M_{\rm BH} = M_d(R_{\rm inf}) + M_{\rm 
halo}(R_{\rm inf})$, is indicated by the vertical dashed line on each panel. 
For the 
three galaxy masses considered, $R_{\rm inf}$ ranges from 0.05 to 0.76 kpc, 
corresponding to angular diameter sizes, $\theta = (2 R_{\inf}/D_A)$ of 
$0.01''$, $0.06''$ and $0.26''$ respectively, where $D_A$ is the cosmological 
angular-diameter distance to redshift $z = 6$. The horizontal dashed lines show 
the lowest circular velocities (corresponding to the highest spectral 
resolutions) achievable by current and future generation instruments, including
(i) \textit{ALMA} observing the [CII] transition at $z \sim 6$ with channel 
widths of $\sim  2$ MHz $\approx 2.2$ km/s \citep[e.g., Ref.][]{carniani2017}, (ii) the Giant Magellan Telescope
(GMT)\footnote{https://www.gmto.org/} Integral Field Spectrometer (GMTIFS\footnote{https://www.gmto.org/resources/near-ir-ifu-and-adaptive-optics-imager-gmtifs/}; 
with a spectral resolution of $R \sim 10000$ corresponding to $\sim 30$ km/s), and (iii) the 
James Webb Space Telescope  
(\textit{JWST}\footnote{https://www.jwst.nasa.gov/}; whose Mid-InfraRed 
Instrument \citep[MIRI; e.g. Ref.][]{wright2015} has a resolving power of $R = 
\delta \lambda/\lambda \sim 3500$ corresponding to $\sim$ 85.7 km/s).

{ {The main result of Fig. 
\ref{fig:rotvelcurves} is the red curve, which indicates the influence of the 
black hole on the kinematics of the stellar disc. Although the figure 
illustrates this for the stellar component, it is important to note that 
\textit{any} tracer of the underlying dark matter halo can be used to infer the 
mass of the black hole from its radius of influence.}

\begin{figure}
\begin{center}
\includegraphics[width = 0.8\textwidth]{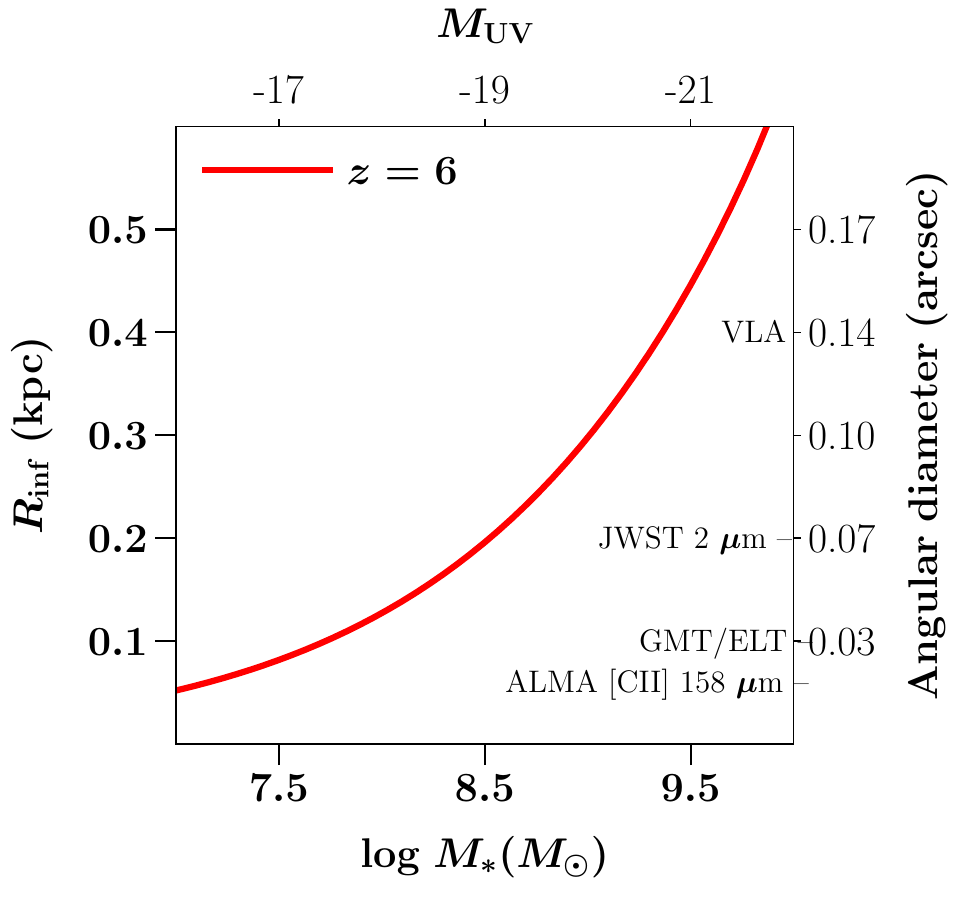} 
\caption{Evolution of the radius of influence, $R_{\rm 
inf}$, of the central black hole as a function of the surrounding stellar mass, 
$M_*$ (lower horizontal axis) and the corresponding absolute magnitude, $M_{\rm 
UV}$ of the host galaxy (upper horizontal axis) at $z = 6$, assuming the same 
dark matter halo, stellar disc and central black hole properties as in 
Fig.\ref{fig:rotvelcurves}. The right-hand  vertical axis indicates the 
corresponding angular diameter scale $\theta = [2R_{\rm inf}/D_A(z = 6)]$, { 
{on 
which the angular resolutions of several upcoming facilities at these redshifts 
(ALMA in its most extended configuration observing [CII] 158 $\mu$m, the 
GMT/ELT, JWST and VLA observing the CO $1-0$) are indicated.}} }
\label{fig:radinfm}
\end{center}
\end{figure}

\begin{figure}
\begin{center}
\includegraphics[width = 0.8\textwidth]{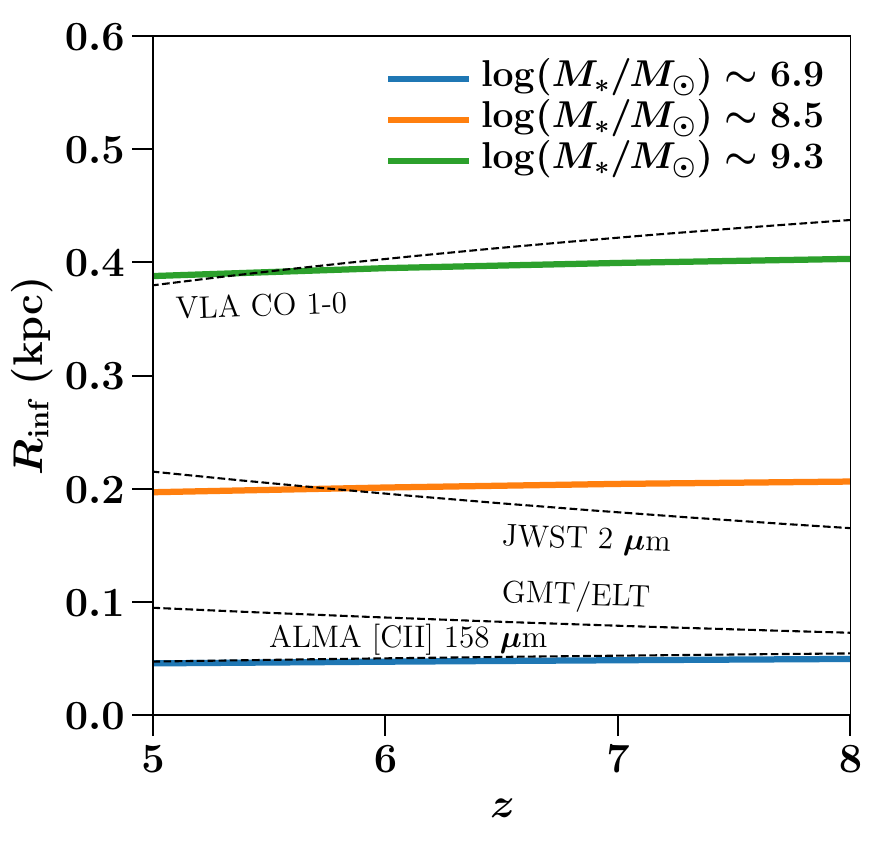}
\caption{Evolution of the black hole radius of influence $R_{\rm 
inf}$ with $z$ for three different halo masses ($\log_{10} (M_{\rm 
halo}/M_{\odot}) = 
10, 11$ and $11.5$ respectively, corresponding to the inferred stellar mass 
range $\log_{10} (M_*/M_{\odot}) \approx 6.9 - 9.3$). { {The angular 
resolutions 
achievable by several current and future facilities (ALMA in its most extended 
configuration observing [CII] 158 $\mu$m, the GMT/ELT, JWST and VLA observing 
the CO $1-0$) are overplotted as dashed lines.}}}
\label{fig:radinfz}
\end{center}
\end{figure}

{ {The evolution of $R_{\rm inf}$ with stellar mass is shown in Fig. 
\ref{fig:radinfm}. The figure shows that $R_{\rm inf}$ evolves strongly with 
stellar masses in the range $10^7 - 10^{10} M_{\odot}$ at $z \sim 6$, as can 
also be seen from Fig. \ref{fig:rotvelcurves}. The upper $x-$axis shows the 
absolute magnitudes ($M_{\rm 
UV}$) corresponding to this mass range, computed from the relation of
 Ref. \citep{tacchella2018} at $z = 6$.
 
 Fig. \ref{fig:radinfz} shows $R_{\rm inf}$ at three different halo masses 
($\log_{10} (M_{\rm 
halo}/M_{\odot}) = 10, 11, 11.5$)  across the range $z \sim 5-8$. This halo 
mass 
range corresponds to a stellar mass range of $\log_{10} (M_*/M_{\odot}) \approx 
6.9 - 9.3$, which is accessible to HST at these redshifts \citep{holwerda2015}. 
The near-constant black hole - stellar mass relation in this $z$-range implies 
that $R_{\rm inf}$ does not evolve much with redshift for a given halo mass.

For the highest stellar masses, the black hole can dominate the kinematics up 
to 
a distance $> 0.5$ kpc, which corresponds to an angular scale of $> 0.2''$. 
These values are well within the reach of current and upcoming obervations, the 
angular resolutions of some of which are indicated on Figs. \ref{fig:radinfm} 
and \ref{fig:radinfz}.  Many observatories in the next generation of 
ground-based telescopes 
(GMT, as well as ELT\footnote{https://www.eso.org/sci/facilities/eelt/}, 
and TMT\footnote{https://www.tmt.org/}) 
will 
have adequate spatial resolution ($\lesssim 30$ mas) to search for these black 
holes.
$ALMA$ in its extended 12-m configuration should be able to detect 
[CII]-emitting galaxies with 0.02 - 0.043$''$ resolution at frequencies 
corresponding to $z \sim 6-10$, and the
$JWST$ should be able 
\citep{vogelsberger2019} to resolve at least $\sim 200$ 
such galaxies at $z \sim 8$,   {with a resolution of $\sim 68$ mas at $2 \ \mu 
m$\footnote{https://sci.esa.int/web/jwst/-/47519-gardner-j-et-al-2009}.} 
Further, the 
$VLA$\footnote{https://science.nrao.edu/facilities/vla/} and the Phase II of 
the 
planned Square Kilometre Array \citep[SKA;][]{taylor2013} will be able to 
access 
the CO J $1\to0$ transition at redshifts $z \sim 6-10$ with angular resolutions 
of 0.2-0.3$''$, which could potentially resolve $R_{\rm inf}$ of the central 
black hole in the brightest galaxies at these epochs.
}}

\section{Conclusions}

These predictions have far-reaching implications. First, a direct 
\textit{kinematic} measurement of the central black hole masses will enable key 
physics insights into the growth and merger histories of supermassive black 
holes in the earliest galaxies. It would also shed light on the active phases 
(`duty cycles') of the earliest black holes, and permit exciting constraints on 
the currently debated contribution of quasars to cosmic reionization of 
hydrogen 
and helium \citep{madau2015}. It has been recently suggested that a significant 
population of strongly lensed quasars \citep{fan2019, wyithe2002lensed} in the 
early universe may have been missed by current surveys due to selection effects 
that rule out bright lenses \citep{pacucci2019}. Direct black hole mass 
measurements form a complementary dataset to aid detections of lensed quasars 
from future surveys.
  Finally, a characterization of the black hole mass function will enable 
forecasting of future measurements of gravitational radiation emitted by pairs 
of these produced by galaxy mergers \citep{Goulding2019}, in anticipation of 
the 
upcoming Laser 
Interferometer Space Antenna ($LISA$\footnote{https://www.lisamission.org}).
  
\section*{Acknowledgements}  The work of AL was supported in part by Harvard's 
Black Hole Initiative, which is funded by grants from JTF and GBMF. We thank 
the anonymous referee for a detailed and helpful report that improved the 
content and the quality of presentation.

\def\aj{AJ}                   
\def\araa{ARA\&A}             
\def\apj{ApJ}                 
\def\apjl{ApJ}                
\def\apjs{ApJS}               
\def\ao{Appl.Optics}          
\def\apss{Ap\&SS}             
\def\aap{A\&A}                
\def\aapr{A\&A~Rev.}          
\def\aaps{A\&AS}              
\def\azh{AZh}                 
\def\baas{BAAS}
\def\jcap{JCAP}
\def\jrasc{JRASC}             
\def\memras{MmRAS}
\def\na{New Astronomy}
\def\nat{Nature}
\def\mnras{MNRAS}             
\def\pra{Phys.Rev.A}          
\def\prb{Phys.Rev.B}          
\def\prc{Phys.Rev.C}          
\def\prd{Phys.Rev.D}          
\def\prl{Phys.Rev.Lett}       
\def\pasp{PASP}               
\def\pasj{PASJ}
\def\physrep{Phys. Repts.}
\def\qjras{QJRAS}             
\def\skytel{S\&T}             
\def\solphys{Solar~Phys.}     
\def\sovast{Soviet~Ast.}      
\def\ssr{Space~Sci.Rev.}      
\def\zap{ZAp}                 
\let\astap=\aap
\let\apjlett=\apjl
\let\apjsupp=\apjs

\bibliography{mybib}{}
\bibliographystyle{JHEP}

\end{document}